\documentclass{elsart}

\usepackage{graphics}
\usepackage{graphicx}

\usepackage{amssymb}

\newbox\grsign \setbox\grsign=\hbox{$>$} \newdimen\grdimen \grdimen=\ht\grsign
\newbox\simlessbox \newbox\simgreatbox
\setbox\simgreatbox=\hbox{\raise.5ex\hbox{$>$}\llap
     {\lower.5ex\hbox{$\sim$}}}\ht1=\grdimen\dp1=0pt
\setbox\simlessbox=\hbox{\raise.5ex\hbox{$<$}\llap
     {\lower.5ex\hbox{$\sim$}}}\ht2=\grdimen\dp2=0pt
\def\simgreat{\mathrel{\copy\simgreatbox}}

\begin{document}

\begin{frontmatter}



\title{Nonextensivity and Galaxy Clustering in the Universe}


\author[das]{C.A. Wuensche}
\author[uesc]{A.L.B.Ribeiro}
\author[lac]{F.M.Ramos}
\author[lac]{R.R.Rosa}

\address[das]{DAS, Instituto Nacional de Pesquisas Espaciais}
\address[uesc]{DCET, Universidade Estadual de Santa Cruz}
\address[lac]{LAC, Instituto Nacional de Pesquisas Espaciais}

\begin{abstract}
We investigate two important questions about the
use of the nonextensive thermostatistics (NETS) formalism in the context
of nonlinear galaxy clustering in the Universe. Firstly, we
define a quantitative criterion for justifying nonextensivity
at different physical scales. Then, we discuss the physics behind
the ansatz of the entropic parameter $q(r)$. Our results suggest
the approximate range where nonextensivity can be justified and, hence,  give
some support to the applicability of NETS to the study of large scale structures.
\end{abstract}

\begin{keyword}
nonextensivity \sep large scale structure \sep multiscaling
\PACS 
\end{keyword}
\end{frontmatter}

\section{Introduction}
\label{1}

The evolution of large scale structures in the Universe is
one of the most important questions of modern cosmology.
A considerable amount of work has been done on numerical and
analytical approaches to characterise the clustering of matter at large
scales. One of the difficulties inherent to this subject is that, 
in order to make progress in the understanding of general models, it is necessary
to define methods for structure quantification.

Usually, quantitative
studies of large scale structures
are based on the two-point correlation function $\xi(r)$
for the galaxy distribution. Estimates indicate this function is
well approximated by $\xi(r)=(r/r_0)^{-\gamma}$ (with $\gamma\approx 1.8$), where
$r_0$ is the correlation length, the scale marking the transition
between linear and non-linear regimes. The usual range for $\gamma$, found in the 
literature, is $1.5 < \gamma < 1.97$ (see, e.g., \cite{hawkins} \cite{snethlage}),
with $r_0 = 5{\rm h}^{-1}$ Mpc for galaxies and $r_0 = 25{\rm h}^{-1}$ Mpc 
for cluster (see, e.g., \cite{peebles}).
At larger scales, the structure of the Universe presents patterns like
walls and filaments (with dimensions $\sim 150 {\rm h^{-1}}$ Mpc) seeemgly reaching
homogeneization at Cosmic Microwave Background (CMB) scales ($\simgreat 1000{\rm h^{-1}}$ Mpc). 
Some authors have had some success describing
the clustering properties of visible matter over this wide
range of scales in terms of a multifractal phenomenon
associated with density thresholds applied to multifractal sets (e.g. 
\cite{martinez}
\cite{pan}).
However, the relative success of the multifractal approach does not imply a
better understanding of the physics behind this framework. Actually, it is
not simple to find a dynamical connection between fractal sets and
galaxy clustering. Recently, Ramos et al. \cite{ramos1} (hereafter RWRR) 
shed some light on this question by putting forward a model based on the generalized
thermodynamics (NETS) formalism \cite{tsallis}. They show
that applying the idea of nonextensivity, intrinsic to NETS, it is
possible to derive an expression for the correlation function

\begin{equation}
1+\xi(r)={1\over 3} D_2 r^{(D_2-3)},
\end{equation}

\noindent using a scale-dependent correlation dimension

\begin{equation}
D_2(r) = 3{{\rm log}~ [2+a(1-q(r))]\over {\rm log}~ 2},
\end{equation}

\noindent where the entropic parameter $q(r)$ is given
by the following ansatz

\begin{equation}
\label{qq}
r\sim {1\over (q-1)^\beta}
\end{equation}

\noindent with $a$ and $\beta$ being free parameters of the model. This approach
shows a smooth transition from a clustered Universe to
large-scale homogeneity, with $D_2=3$. However, RWRR
do not discuss two important questions concerning the conceptual
basis of the model: the necessity of defining a criterion which allows us to assume nonextensivity in
the context of galaxy clustering and the physics behind the ansatz for $q(r)$. 
In the next two sections these questions are
further developed and some conclusions are drawn at the end.

\section{Nonextensivity and gravitational clustering}
\label{2}

The physical motivation of the approach adopted by RWRR is built upon the fact that 
components of gravitating systems tend to evolve 
spontaneously into increasingly complex structures due to the long range nature 
of the gravitational interaction. 
The NETS theory generalises the Boltzmann-Gibbs statistical mechanics and
can be applied to systems dominated by the long-range nature of gravity.
However, the application of NETS to an ensemble of comoving cells containing
gravitating particles depends on the behaviour of the average correlation
energy inside a spherical cell of volume $V=4\pi R^3/3$ with increasing scales $R$.
For instance, the grand canonical ensemble of cells that are larger than the
correlation length  are approximately extensive, since the universal
expansion effectively limits the thermodynamic effects of gravity to roughly
the correlation length scale (e.g. 
\cite{peebles}
\cite{safang}).
Also, for an individual cell whose size is larger than the correlation length,
extensivity is possibly a good approximation because the correlation energy between two members
of the ensemble is negligible compared to the internal correlation energy:

\begin{equation}
U_{corr} \ll U_1 + U_2 ~~\Rightarrow U_{tot} \approx U_1 + U_2.
\end{equation}

Due to these reasons, the application of NETS to galaxy clustering is not
straightforward and demands better criteria to properly describe the problem
using the non-extensive formalism. For real nonextensive systems, the correlation energy between two
cells should be as important as the internal correlation energy, such that

\begin{equation}
U_{tot} = U_1 + U_2 + U_{corr}.
\end{equation}

\noindent Following
Sheth \& Saslaw \cite{sheth}, we compute the average gravitational correlation
energy within cells of volumes $V$ and $2V$ for increasing $R$:

\begin{equation}
W_V = \bar{n}V\int_0^R {Gm^2\over 2r}\xi(r)4\pi\bar{n}r^2\;dr,
\end{equation}

\noindent where $m$ is the mass and $\bar{n}$ is the average number density of particles in a
cell of size $V$. The extensivity approximation requires that $W_{2V}\approx 2W_V$,
approximately verified for the  power law correlation,  $\xi(r)=\xi_0r^{-\gamma}$,
whenever $\gamma\geq 1$ (see \cite{sheth}). 
In this case, $|W_{2V}/2W_{V}|=2^{(2-\gamma)/3}$, which means that, using $\gamma=1.77$, 
we find a $\sim$5\% deviation from
the strict extensivity condition at all scales.

In the NETS context, we substitute (1), (2) and (3) into (6), numerically integrate for $V$ and $2V$
(the spherical cell of volume $2V$ has radius $2^{1/3}R$) and calculate the correlation
ratio $|W_{2V}/2W_{V}|$ for each upper limit $R$. The results are presented in Figure 1, 
where we plot the three cases investigated by RWRR  in comparison to the line defined by 
the power law correlation (for $\gamma=1.77$).
Note that, for the three NETS models, we
see different levels of deviations from the extensivity approximation. 
In particular, ${\rm NETS_-3}$ model presents
deviations from extensivity of $\sim$20\% even at very large scales. However,
for large enough scales,
the correlation ratio decreases and the extensivity approximation is recovered,
for models ${\rm NETS_-1}$ and ${\rm NETS_-2}$. In order to properly 
invoke nonextensivity, we propose
a lower limit for the ratio $|W_{2V}/2W_{V}|$ as 10\%. This is twice the
energy ratio for the power law correlation case. 

\begin{figure}[h]
  \center{
    \includegraphics[width=10cm,height=8cm]{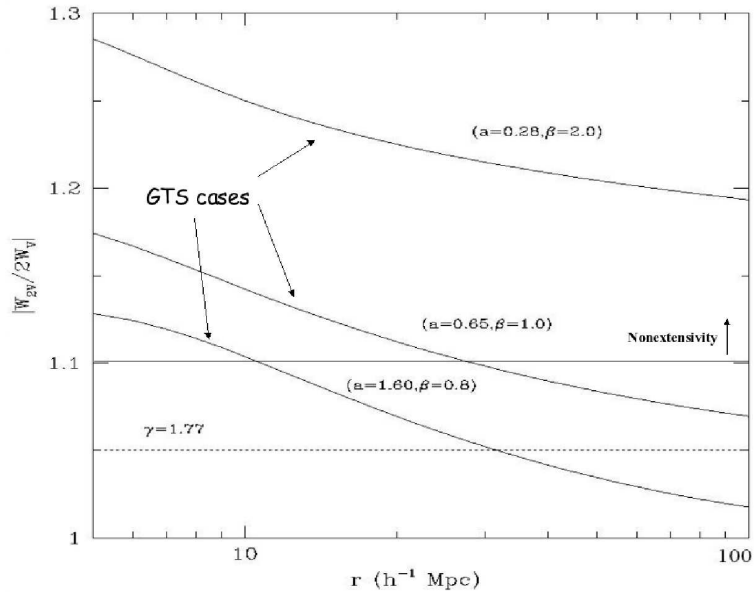}
}
   \caption{\label{Fig 1}Behaviour of the correlation energy ratio with the scale $R$.}
\end{figure}

One can further quantify that as
follows. Consider two elliptical galaxies of radii $r_g$ separated 
by the distance $r_0$. A family of analytic models for spheroidal stellar systems
is defined by the density distribution:

\begin{equation}
\rho_\eta(r) ={\eta\over 4\pi}{1\over r^{3-\eta}(1+r)^{1+\eta}}
\end{equation}

\noindent for $0<\eta \le 3$ \cite{tremaine}. Choosing units in which the
total mass  and the gravitational constant are both unity, we find the self-gravitational
energy of each galaxy as 

\begin{equation}
U_1=U_2 = -{1\over 2}\left({1\over 2\eta -1}\right)=U_\eta,
\end{equation}

\noindent and the potential at a distance $r_0$ is

\begin{equation}
\Phi_\eta (r_0) = {1\over \eta -1}\left[{r_0^{\eta-1}\over (1+r_0)^{\eta-1}} -1\right],
\end{equation}

\noindent with the mass interior to radius $r_g$ being

\begin{equation}
M_\eta (r_g) = {r_g^\eta\over (1+r_g)^\eta}.
\end{equation}

\noindent The gravitational energy of the interacting system formed by
the two galaxis is given by 

\begin{equation}
U_{int} = 2\Phi_\eta (r_0) M_{\eta} (r_g).
\end{equation}

\noindent Assuming we have only the two galaxies in the volume defined by $4\pi r_0^3/3$,
we should have $U_{corr}=U_{int}$. Then $U_{tot}=2U_\eta + U_{int}$ and, consequently,

\begin{equation}
{U_{tot}\over 2U_\eta} = 1-2\left({2\eta-1\over\eta-1}\right)\left[{r_0^{\eta-1}\over
(1+r_0)^{\eta-1}} -1\right]\left[{r_g^\eta\over (1+r_g)^\eta}\right].
\end{equation}

\noindent In Figure 2 we present the behaviour of ${U_{tot}/ 2U_\eta}$ as
a function of $r_0$ (for $r_0<5{\rm h}^{-1}$ Mpc and taking 
$r_g=30{\rm h}^{-1}$ kpc).
Note that only at very small scales the ratio significantly increases.
Actually, at the galaxy correlation length, the deviation from extensivity
is about 1\%. As a comparison, NETS models give $\sim$ 13\%, 17\% and 28\%,
for the cases 1, 2 and 3,  respectively, at the same scale, in the more general
situation. Thus, a NETS model with
at least 10\% of deviation from the strict extensivity condition
corresponds to 10 times the deviation in the case of a cell with
only two typical elliptical galaxies interacting within the galaxy
correlation length ($r_0=5{\rm h}^{-1}$).

Using this criterium, ${\rm NETS_-1}$ and ${\rm NETS_-2}$ models are
well justified for $r< 10$ ${\rm h^{-1}}$  Mpc and  $r< 30$ ${\rm h^{-1}}$ Mpc,
respectively, while  ${\rm NETS_-3}$ model is valid at all scales.
This result reinforces the idea of applying the NETS formalism to galaxy
clustering phenomena, but clearly indicating the approximate range 
it will be used, given a specific choice of the free parameters
$a$ and $\beta$. Hence,
the level of nonextensivity strongly depends on the the ansatz (3), 
which reinforces the need of a deeper discussion of the meaning of $q(r)$.

\begin{figure}
  \center{
     \includegraphics[width=8cm,height=12cm,angle=270]{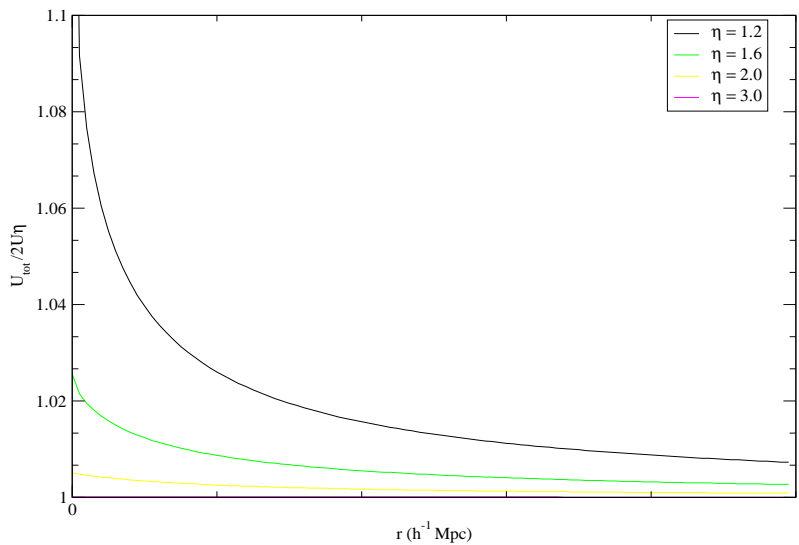}
  }
  \caption{\label{Fig 2} Correlation energy ratio for $\eta$ models. For comparison, 
the Hernquist's model is computed for $\eta=2$.}
\end{figure}

\section{The physics behind q(r)}
\label{3}

Structure formation models try to compute the evolution of cosmic structure from 
the very early Universe to the present day.
Usually, they are stochastic, in the sense that random initial conditions are 
used, with well-specified statistical properties, and their late evolution may 
be highly non-linear 
\cite{bertschinger}. Any attempt to compute $q(r)$ from 
first principles must take into account this complex scenario.

Considering the relevant role played by turbulence in the dynamics of structure
formation (e.g. \cite{devega} \cite{larson}), 
a natural approach for deriving a 
meaningful expression for $q(r)$
is to assume a {\it{a la}} Kolmogorov cascade phenomenology, in which the 
energy supplied at large scales by the physical mechanisms acting over
structure formation  flows down until
being finally dissipated on the smallest scales by viscous processes 
\cite{fleck}. The key 
ingredient in this ``top-down" cascade scenario is the presumption of the existence,
within a certain range of scales, of a scaling $\langle v^n_r \rangle \sim r^{\zeta_n}$ 
of the moments of the peculiar velocity differences
$v_r(x) = v(x+r) - v(x)$.  
This scale-invariance can be rigorously deduced, under assumptions such as local isotropy,
from the dynamical equations governing a turbulent fluid \cite{monin}. 
In the present cosmological context, scale-invariance is well supported by 
observations \cite{borgani} and is at the heart of the fractal description 
of galaxy clustering \cite{martinez}. If we now assume that the probability 
density function (PDF) of peculiar velocity differences is described
 by the Tsallis canonical distribution and 
that at sufficiently large scales turbulent fluctuations are normally 
distributed, then an analytical expression for $q(r)$ can be easily 
derived \cite{ramos2}, in the form of $q \sim (15 - 21 r^{\alpha})~/~(9 - 15 r^{\alpha})$,
where $\alpha = \zeta_4 - 2 \zeta_2$ and $\zeta_p$ are the structure function exponents.

This model was successfully applied to hydrodynamics turbulence (\cite{bolzan},\cite{ramos3}),
where the entropic parameter represents a direct measure of intermittency. 
It gives similar results to those obtained with equation (\ref{qq}), 
albeit only within a restricted range of scales. Note that, as $r \rightarrow 0$, the entropic 
parameter tends to a finite value, which is physically expected  
for any model of $q(r)$ \cite{castagnoli}. But in this case, equation (\ref{qq}) 
would not provide a smooth transition from small-scale fractality to large-scale homogeneity, 
which represents a major drawback in the present cosmological context.

The problem with such a cascade phenomenology is its 
description of structure formation from larger to smaller scales. 
To support observations, a different and
more compelling  model may be obtained from a ``bottom-up" fractal 
cascading scenario. In this cold dark matter (CDM) dominated Universe scenario, 
large scale structures are formed by gravitational
clustering of smaller clumps of matter \cite{castagnoli}. This process
eventually leads to strong density fluctuations with self similar density 
distribution and a stationary fractal dimension, while enhancing 
long-range correlations and the corresponding high energy tails 
in the peculiar velocity differences PDFs. 

In this context, averaging a Gaussian conditional velocity 
distribution over all possible 
spatio-temporal energy dissipation rate fluctuations and at appropriate scales leads
to a Tsallis peculiar velocity differences PDF and to a closed-form expression for 
the entropic parameter \cite{beck1}. This expression depends on 
the (discrete) number of degrees of freedom relevant to represent the local fluctuations.
It can be further generalized, taking the form $q \sim (r + 3)~/~(r + 1)$.
Considering the lack of observational data for an unambiguous 
determination of $q(r)$, we foresse a complementary (and computationally expensive) 
approach to gain a deeper insight on the role of the entropic parameter. We propose
to compute estimates of $q$ within a large range of scales directly from $N$-body 
simulations, using a $\Lambda-CDM$ cosmological 
model of gravitational clustering.

\section{Summary and conclusions}
\label{4}

The basis for clustering statistics in cosmology is the study of
galaxy distribution in the Universe. The analysis of an increasingly amount of 
astronomical data can
provide the correct framework to explain the formation and evolution
of large-scale structure in the Universe. In particular, correlation
functions works suggest the possibility of describing the galaxy
clustering properties in the context of the multifractal approach.
Recently, RWRR presented a better physical interpretation
to this approach by deriving multifractality from the NETS formalism.
Now, extending their work, we define
a quantitative criterium to use the NETS formalism based on deviations
 from strict extensivity. Assuming it should reach
at least 10 times the expected deviation level within a cell containing
only two typical galaxies, we obtain the approximate validity domain for the NETS models:
$r <10$ Mpc for ${\rm NETS}_{-}1$ model, $r < 30$ Mpc for  
${\rm NETS}_{-}2$, and at any scale for  ${\rm NETS}_{-}3$.
The choice of $q(r)$ has
a physical meaning in the context of hydrodynamics turbulence and
can be understood as a ``bottom-up'' fractal cascading,  in conceptual agreement
with CDM structure formation models. Although the scenario we propose
is not complete, we should keep in mind that there is not a unified 
framework to explain galaxy clustering in the Universe. Further numerical developments 
of NETS models will be presented in a forthcoming paper \cite{dantas}.

This work was partially supported by FAPESP grant 00/06770-2. 
FMR also thanks the support given by CNPq through the research grant 300171/97-8. 
CAW thanks the partial support of CNPq through the research grant 300409/97-4-FA.



\end{document}